\documentclass{article}
\usepackage{spconf}      
\usepackage{amsmath}     
\usepackage{amssymb}     
\usepackage{amsfonts}    
\usepackage{graphicx}    
\usepackage{caption}     
\usepackage{tabularx} 
\usepackage{multirow}

\usepackage{algorithm}
\usepackage{algpseudocode}
\usepackage{booktabs}

\usepackage{hyperref}

\title{DDSC: Dynamic Dual-Signal Curriculum for Data-Efficient Acoustic Scene Classification under Domain Shift}
\name{Peihong Zhang, Yuxuan Liu, Rui Sang, Zhixin Li, Yiqiang Cai, Yizhou Tan, Shengchen Li}
\address{School of Advanced Technology, Xi’an Jiaotong-Liverpool University, Suzhou, China}
\ninept
\begin{document}
%
\maketitle
\begin{abstract}

Acoustic scene classification (ASC) suffers from device-induced domain shift, especially when labels are limited. Prior work focuses on curriculum-based training schedules that structure data presentation by ordering or reweighting training examples from easy-to-hard to facilitate learning; however, existing curricula are static, fixing the ordering or the weights before training and ignoring that example difficulty and marginal utility evolve with the learned representation. To overcome this limitation, we propose the Dynamic Dual-Signal Curriculum (DDSC), a training schedule that adapts the curriculum online by combining two signals computed each epoch: a domain-invariance signal and a learning-progress signal. A time-varying scheduler fuses these signals into per-example weights that prioritize domain-invariant examples in early epochs and progressively emphasize device-specific cases. DDSC is lightweight, architecture-agnostic, and introduces no additional inference overhead. Under the official DCASE 2024 Task~1 protocol, DDSC consistently improves cross-device performance across diverse ASC baselines and label budgets, with the largest gains on unseen-device splits.

\end{abstract}
\section{Introduction}
\label{sec:intro}

\begin{figure}[t]
    \centering
    \includegraphics[width=0.5\textwidth]{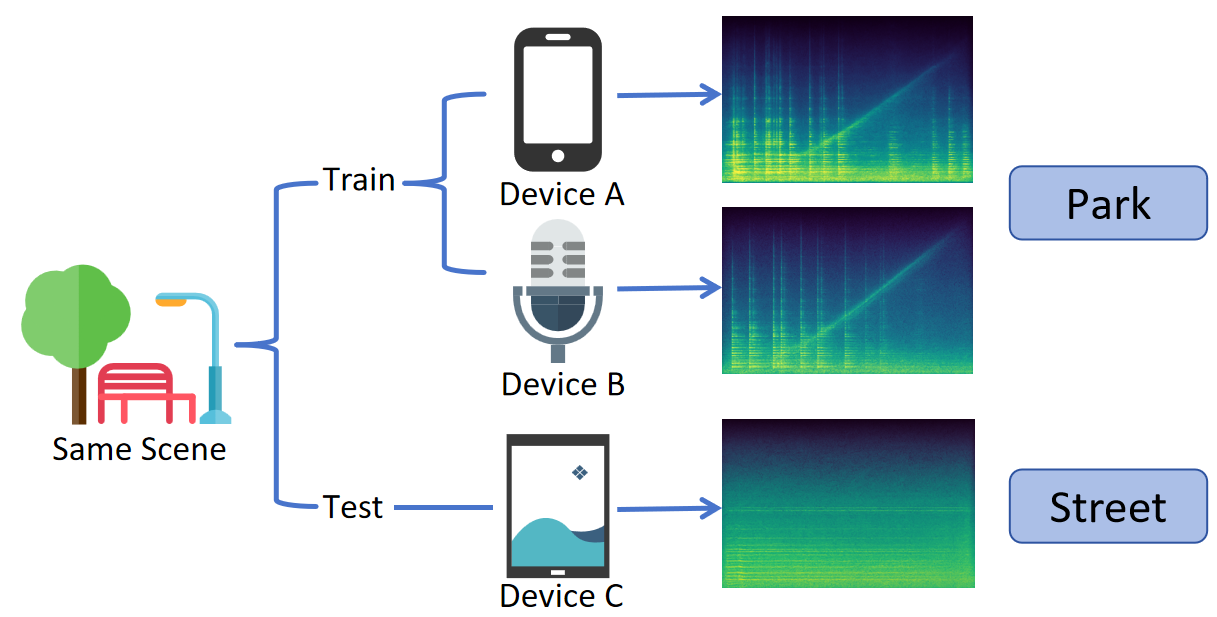} 
    \caption{Device-induced domain shift in ASC. For the same scene, recordings captured with different devices exhibit distinct spectrograms due to device characteristics. Consequently, a model trained on devices A and B may misclassify inputs from an unseen device C, highlighting the challenge of unseen-device generalization.}

    \label{fig:domain_shift}
\end{figure}

Acoustic scene classification (ASC) aims to recognize the scene from an audio segment and is a fundamental task in machine hearing \cite{barchiesi2015acoustic}. A central challenge in ASC is domain shift \cite{bai2024description} caused by recording-device variation, which may degrade generalization to unseen devices \cite{tan2024acoustic}, shown in Fig.~\ref{fig:domain_shift}. The issue is exacerbated when limited labeled data are available \cite{Zhang2025}. The DCASE 2024 Challenge Task 1 \cite{schmid2024data} explicitly targets this scenario by requiring systems to be trained on small fractions (as low as 5\%) of labeled audio from a limited set of devices and evaluated on recordings from unseen devices, while specifying strict limits on model size and computation.


To mitigate domain shift with limited annotations, prior work \cite{Zhang2025,wang2024curriculum,wang2024ladder,zhao2024symmetric} explores Curriculum Learning (CL) to encourage domain-invariant representations. CL mimics the human progression from easy to hard and structures the presentation of training data \cite{bengio2009curriculum}, facilitating the acquisition of domain-invariant, generalizable representations \cite{liu2022less,zhang2025nmcse}.Following this paradigm, these methods predefine an easy-to-hard curriculum that reorders or reweights training examples, encouraging the model to first focus on domain-invariant samples and later incorporate device-specific ones.



However, existing curricula exhibit a critical limitation: they are static—the ordering or weights are fixed prior to training and remain unchanged thereafter \cite{Zhang2025,wang2024curriculum,wang2024ladder}. This design neglects that both per-example difficulty and the marginal utility of each example for learning device-invariant structure change as the learned representation and the decision boundary evolve during training \cite{zhao2024symmetric}. The result is a skewed exposure pattern: samples judged as high-invariance at early stages are overemphasized, whereas difficult cross-device cases—those informative for refining domain boundaries later—receive insufficient attention \cite{wang2021survey}. We contend that sample difficulty is nonstationary: as relevant knowledge is acquired, previously difficult examples can become easier, whereas initially easy examples saturate and contribute diminishing signal. These considerations motivate a dynamic curriculum that updates per-example weights online during training.

Accordingly, a central design question arises: how should one design online, model-coupled signals to assess per-example difficulty as training evolves? Such signals should jointly encode (i) domain invariance—robustness across devices—and (ii) learning contribution—measuring the marginal benefit of revisiting a sample.

To this end, we propose the Dynamic Dual-Signal Curriculum (DDSC), a dynamic curriculum that adaptively adjusts sample weights. At each epoch, DDSC computes two complementary signals: (i) a domain-invariance score obtained from prototype–posterior entropy over device prototypes, and (ii) a learning-progress score derived from the smoothed change of the per-example loss across epochs. A time-varying scheduler fuses these signals into a continuous curriculum score that determines per-example weights. The resulting schedule realizes an easy-to-hard presentation—prioritizing domain-invariant examples in early epochs and progressively emphasizing device-specific, difficult cases—thereby improving generalization to unseen devices without architectural changes or inference overhead.


To assess the effectiveness of DDSC, we evaluate it on the DCASE 2024 Task 1 dataset \cite{heittola2020acoustic} using the official splits across multiple ASC systems with diverse architectures and training setups \cite{schmid2023distilling,cai2024dcase2024,han2024data,chen2025improving}. Experimental results show that DDSC consistently improves cross-device performance under low-label budgets and outperforms existing curriculum baselines \cite{Zhang2025,zhao2024symmetric,wang2024curriculum,wang2024ladder}, particularly on unseen-device splits. By design, DDSC is lightweight, architecture-agnostic, and easily integrable—it requires no modifications to the model architecture and introduces no additional inference overhead. As a training-schedule method, DDSC complements data augmentation \cite{schmid2022cp,cai2023dcase2023} and feature-alignment approaches \cite{zhao2022feature} for device-induced domain shift, providing a practical tool for domain generalization in data-efficient ASC.

\section{Dynamic Dual-Signal Curriculum}

\begin{figure}[!t]
  \centering
  \includegraphics[width=0.5\textwidth]{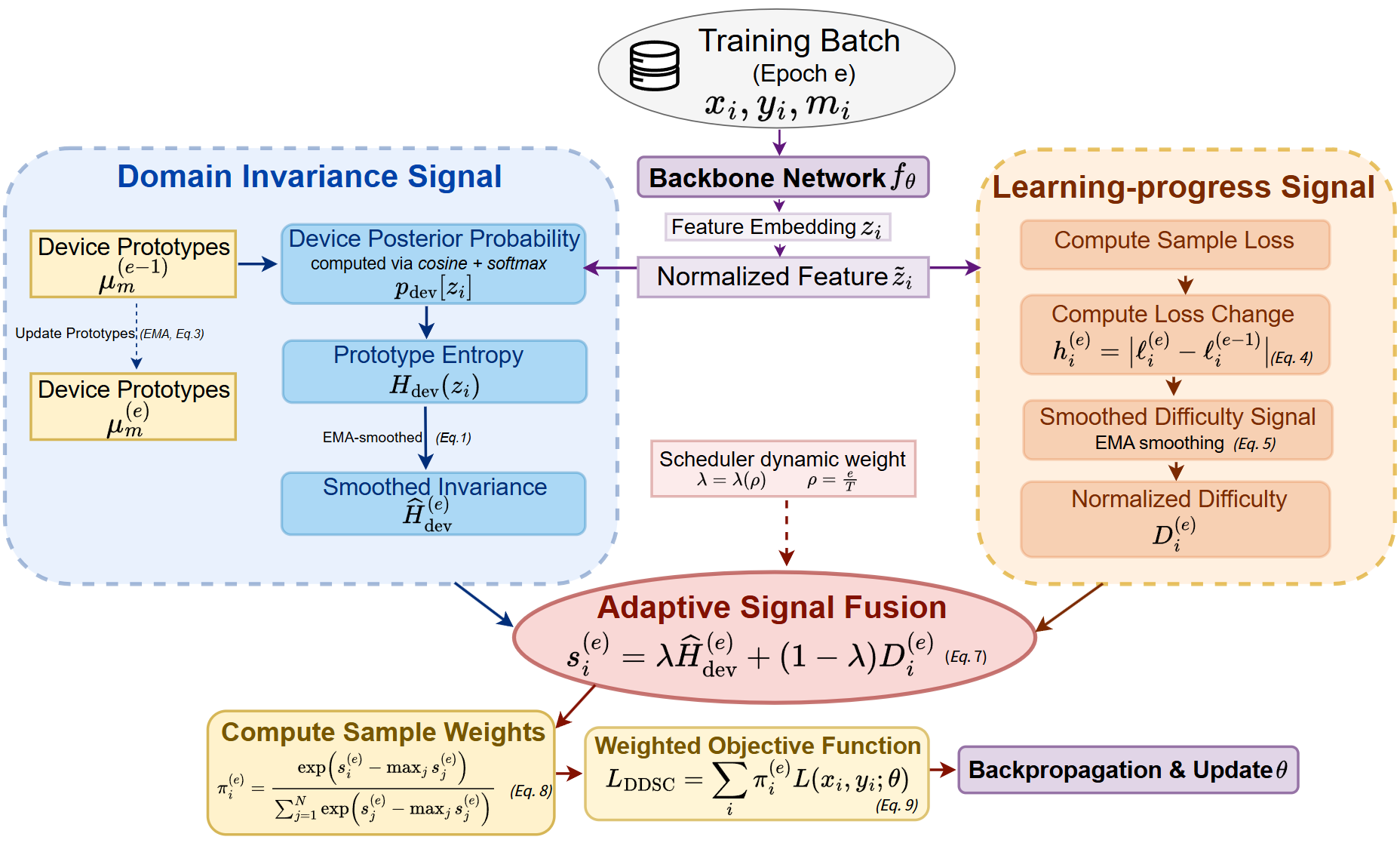}
  \caption{Overview of the DDSC. For each epoch, the backbone produces normalized features; the domain-invariance and learning-progress signals are computed and fused by a time-varying scheduler to produce per-example weights used in the weighted objective.}
  \label{fig:ddsc-overview}
\end{figure}

\subsection{Notation and Objective}
\label{subsec:notation}

Let $\mathcal{D}_{\mathrm{train}}=\{(x_i,y_i,m_i)\}_{i=1}^{N}$, where $x_i$ is an audio clip, $y_i$ its scene label, and $m_i\in\{1,\ldots,M\}$ indexes the recording device among $M$ devices.
A backbone $f_\theta$ maps inputs to features $z_i=f_\theta(x_i)\in\mathbb{R}^{F}$.
Let $L(x_i,y_i;\theta)$ denote the baseline per-sample loss; we do not assume any specific form.

At training epoch $e$, DDSC computes two signals for each example: a domain-invariance signal derived from prototype entropy (Section~\ref{subsec:proto-entropy}) and a learning-progress signal based on a smoothed change of the per-sample loss (Section~\ref{subsec:dih}).
A scheduler (Section~\ref{subsec:fusion}) fuses these signals into a curriculum score $s_i^{(e)}$ and maps scores to nonnegative per-example weights $\pi_i^{(e)}$ with $\sum_{i=1}^{N}\pi_i^{(e)}=1$. DDSC leaves the baseline loss function unchanged and replaces uniform averaging with dynamically computed per-example weights, yielding the epoch-wise weighted objective; pseudocode is shown in Algorithm~\ref{alg:ddsc}, and Fig.~\ref{fig:ddsc-overview} illustrates the method.


\subsection{Domain-Invariance Signal: Prototype Entropy}
\label{subsec:proto-entropy}

Prior work trains a separate device classifier offline and uses its predictive entropy as this indicator—high entropy means the device is hard to identify, thus the sample is more domain-invariant \cite{Zhang2025}. Because that classifier is independent of the ASC model, it does not update with the backbone during training; its outputs are one-shot labels, so the curriculum remains static. To obtain a dynamic curriculum, we use prototype entropy \cite{mavridis2022risk}: with unit-normalized embeddings and device prototypes updated online from detached features, we form a temperature-scaled posterior over devices via cosine similarity and take its entropy as the indicator. This signal co-evolves with the backbone and adds no extra trainable parameters.

\noindent\textbf{Device prototypes and posterior.}
For each device $m\in\{1,\ldots,M\}$ we maintain a unit prototype $\mu_m\in\mathbb{R}^{F}$ estimated online from detached embeddings. 
Given a unit feature $\tilde z=z/\lVert z\rVert_2$, we compute cosine scores and a temperature-scaled posterior
\begin{equation}
\label{eq:cos-score}
s_m(\tilde z)=\tilde z^\top \mu_m,\qquad
p(m\mid \tilde z)=\frac{\exp\!\big(s_m(\tilde z)/\tau\big)}{\sum_{m'=1}^{M}\exp\!\big(s_{m'}(\tilde z)/\tau\big)}.
\end{equation}
The device prototype entropy and its normalized form are
\begin{equation}\label{eq:proto-entropy}
\begin{aligned}
H_{\text{dev}}(\tilde z) &= -\sum_{m=1}^{M} p(m\mid \tilde z)\log p(m\mid \tilde z),\\
\widetilde H_{\text{dev}}(\tilde z) &= \frac{H_{\text{dev}}(\tilde z)}{\log M}\in[0,1].
\end{aligned}
\end{equation}
where a larger $\widetilde H_{\text{dev}}$ indicates weaker device evidence and hence stronger domain invariance.

\medskip\noindent\textbf{Stable prototype estimation.}
Let $\hat z_i = z_i/\|z_i\|_2$ denote the detached unit embedding.
At the end of each training epoch $e$, for each device $m$ collect all unit embeddings observed in that epoch,
$\mathcal{S}_m^{(e)}=\{\hat z_i:\,m_i=m\}$.
If $\mathcal{S}_m^{(e)}=\varnothing$, set $\mu_m^{(e)}=\mu_m^{(e-1)}$.
Otherwise, compute the sample mean
\[
\bar z_m^{(e)}=\frac{1}{|\mathcal{S}_m^{(e)}|}\sum_{\hat z\in\mathcal{S}_m^{(e)}} \hat z,
\]
and update prototypes by an exponential moving average (EMA) followed by $\ell_2$ re-normalization:
\begin{equation}\label{eq:proto-ema}
\tilde\mu_m^{(e)}=(1-\gamma)\,\mu_m^{(e-1)}+\gamma\,\bar z_m^{(e)},\quad
\mu_m^{(e)}=\tilde\mu_m^{(e)}\big/\|\tilde\mu_m^{(e)}\|_2.
\end{equation}
When device $m$ first appears at epoch $e_0$, initialize
\[
\mu_m^{(e_0)}=\bar z_m^{(e_0)}\big/\|\bar z_m^{(e_0)}\|_2.
\]

\subsection{Learning-Progress Signal: Smoothed Loss-Change}
\label{subsec:dih}

At training epoch $e$, let $\ell_i^{(e)}$ be the per-sample loss of $x_i$.
Define the instantaneous loss change
\begin{equation}
\label{eq:dih-instant}
h_i^{(e)}=\big|\ell_i^{(e)}-\ell_i^{(e-1)}\big|,
\end{equation}
and its exponentially smoothed estimate by EMA
\begin{equation}
\label{eq:dih-ema}
D_i^{(e)}=\beta\,D_i^{(e-1)}+(1-\beta)\,h_i^{(e)},\qquad \beta\in(0,1).
\end{equation}
Larger $D_i^{(e)}$ indicates stronger short-horizon learning instability.
For initialization, set $D_i^{(0)}=0$ and $h_i^{(1)}=0$.

\noindent\textbf{Epoch-wise normalization.}
For cross-epoch comparability, we normalize the learning-progress signal each epoch.
Let $D_{\max}^{(e)}=\max_{i} D_{i}^{(e)}$ and $D_{\min}^{(e)}=\min_{i} D_{i}^{(e)}$. We define
\begin{equation}
\overline D_i^{(e)}=\frac{D_i^{(e)}-D_{\min}^{(e)}}{D_{\max}^{(e)}-D_{\min}^{(e)}+\varepsilon}\in[0,1].
\end{equation}
where $\varepsilon$ is a small constant and $\varepsilon>0$.

Smaller $\overline D_i^{(e)}$ signals limited marginal benefit from additional exposure, whereas larger values indicate greater learning potential. Accordingly, the schedule first emphasizes domain-invariant examples and then increases the weight of samples with larger $\overline D_i^{(e)}$, improving cross-device generalization.

\subsection{Signal Fusion and Weighted Objective}
\label{subsec:fusion}

\noindent\textbf{Curriculum score.}
Let $\widehat H_{\text{dev}}^{(e)}(\tilde z_i)=\mathrm{EMA}_{\eta_H}\!\left[\widetilde H_{\text{dev}}(\tilde z_i)\right]$.
We fuse the two signals as
\begin{equation}
s_i^{(e)}=\lambda(\rho_e)\,\widehat H_{\text{dev}}^{(e)}(\tilde z_i)
+\big(1-\lambda(\rho_e)\big)\,\overline D_i^{(e)},
\end{equation}
where $\rho_e=e/T\in[0,1]$ is the training progress and $\lambda(\rho_e)$ decreases monotonically.
This scheduler prioritizes domain-invariant samples in early epochs and gradually shifts emphasis toward samples with higher estimated difficulty as training proceeds, thereby realizing an easy-to-hard exposure under domain shift.

\noindent\textbf{Parameter selection.}
The schedule function $\lambda(\rho_e)$ is defined as a cosine decay (cosine annealing) with a small floor \cite{loshchilov2016sgdr}: 
\[
\lambda(\rho_e)=\lambda_{\min}+(1-\lambda_{\min})\tfrac{1}{2}(1+\cos(\pi\rho_e)),
\]
where $\lambda_{\min}=0.2$ was selected through validation on the DCASE2024 official baseline. This value was found to provide stable performance across all tested models and label budgets.

\noindent\textbf{Continuous weights.}
Scores are mapped to weights via a softmax:
\begin{equation}\label{eq:softmax-pi}
\pi_i^{(e)}=\frac{\exp\!\big(s_i^{(e)}-\max_{j}s_j^{(e)}\big)}{\sum_{j=1}^{N}\exp\!\big(s_j^{(e)}-\max_{j}s_j^{(e)}\big)},
\end{equation}
The weights satisfy $\sum_{i=1}^{N}\pi_i^{(e)}=1$.

\noindent\textbf{Weighted objective.}
Training minimizes
\begin{equation}
\label{eq:weighted-loss}
\mathcal{L}_{\mathrm{DDSC}}^{(e)}=\sum_{i=1}^{N}\pi_i^{(e)}\,L(x_i,y_i;\theta).
\end{equation}

\begin{algorithm}[t]
\caption{Dynamic Dual-Signal Curriculum}
\label{alg:ddsc}
\begin{algorithmic}[1]
\Require Training set $\{(x_i,y_i,m_i)\}_{i=1}^N$, epochs $T$
\State Initialize device prototypes $\mu_m^{(0)}$, per-sample losses $\ell_i^{(0)}\!\gets\!0$
\For{$e=1$ to $T$}
  \If{$e=1$}
    \State $\pi^{(e)} \gets \tfrac{1}{N}\,\mathbf{1}$ 
\Comment{Uniform weights at the first epoch}
  \Else
    \For{$i=1$ to $N$}
      \State $s_i^{(e)} \gets \lambda_e\,H(i) + \big(1-\lambda_e\big)\,\overline D(i)$
    \EndFor
    \State $\pi^{(e)} \gets \mathrm{softmax}\!\big(s^{(e)}-\max_j s_j^{(e)}\big)$ \Comment{Eq.~\eqref{eq:softmax-pi}}
  \EndIf

  \State $\mathcal L^{(e)} \gets \sum_{i=1}^{N}\pi_i^{(e)}\,L(x_i,y_i;\theta)$ \Comment{Eq.~\eqref{eq:weighted-loss}}
  \State Update $\theta$ by backprop on $\mathcal L^{(e)}$
\EndFor

  \For{$i=1$ to $N$}
    \State $\ell_i^{(e)} \gets$ Running mean of $L(x_i,y_i;\theta)$ over epoch $e$
    \State $\Delta\ell_i^{(e)} \gets \ell_i^{(e)}-\ell_i^{(e-1)}$;\quad $D_i^{(e)} \gets \mathrm{EMA}_{\beta}\!\big(|\Delta\ell_i^{(e)}|\big)$
  \EndFor
  \State $\mu_m^{(e)} \gets (1-\gamma)\mu_m^{(e-1)}+\gamma\,\bar z_m^{(e)}$ 
  \For{$i=1$ to $N$}
    \State Compute device posterior $p_m(i)$ from $\{\mu_m^{(e)}\}$; $\widetilde H_{\mathrm{dev}}^{(e)}(i)\gets -\sum_m p_m(i)\log p_m(i)$
    \State $\widehat H_{\mathrm{dev}}^{(e)}(i)\gets \mathrm{EMA}_{\eta_H}\!\big[\widetilde H_{\mathrm{dev}}^{(e)}(i)\big]$
    \State $H(i)\gets \widehat H_{\mathrm{dev}}^{(e)}(i)$;\quad $\overline D(i)\gets \overline D_i^{(e)}$
  \EndFor

\end{algorithmic}
\end{algorithm}

\section{Experiment}

\subsection{Experimental Setup}

\subsubsection{Datasets} 
Our experimental evaluation is performed using the DCASE 2024 Task 1 dataset \cite{heittola2020acoustic}, a benchmark derived from the TAU Urban Acoustic Scenes 2022 Mobile development set for the Acoustic Scene Classification (ASC) task. The dataset consists of one-second audio clips representing ten distinct acoustic scenes, recorded across 12 European cities with multiple real and simulated mobile devices. In total, the development set contains 230,350 audio segments, amounting to approximately 64 hours of audio.

As detailed in Table~\ref{tab:dataset_distribution}, the dataset is partitioned into training and testing sets with a specific device distribution. The training set includes labeled data from real devices (A, B, C) and simulated devices (S1, S2, S3). A characteristic of the training data is its imbalance, with Device A contributing a substantial majority of the segments. The test partition comprises data from all training devices and, critically, introduces three previously unseen simulated devices (S4, S5, S6) to assess the generalization capabilities of the models.

\begin{table}[H]
\centering
\caption{Distribution of DCASE 2024 Task~1 Development Dataset}
\label{tab:dataset_distribution}
\setlength{\tabcolsep}{1pt} 
\begin{tabular}{@{}ll|rrr@{}}
\toprule
\textbf{Devices} & \textbf{Type} & \textbf{Total segments} & \textbf{Train segments} & \textbf{Test segments} \\
\midrule
A & Real & 144,000 & 102,150 & 3,300 \\
B & Real & 10,780 & 7,490 & 3,290 \\
C & Real & 10,770 & 7,480 & 3,290 \\
S1, S2, S3 & Simulated & $3\times10,800$ & $3\times7,500$ & $3\times3,300$ \\
S4, S5, S6 & Simulated & $3\times10,800$ & -- & $3\times3,300$ \\
\midrule
\textbf{Total} & & \textbf{230,350} & \textbf{139,620} & \textbf{29,680} \\
\bottomrule
\end{tabular}
\end{table}

To investigate performance under data-limited conditions, we strictly adhere to the official low-resource protocol of the DCASE 2024 Challenge. This protocol provides five predefined training subsets drawn from the multi-device data (A, B, C, S1, S2, S3), which contain 5\%, 10\%, 25\%, 50\%, and 100\% of the total labeled data, respectively. All models in our study are trained exclusively on these specified low-resource subsets.

\begin{table*}[t]
\centering
\scriptsize
\setlength{\tabcolsep}{6pt}
\renewcommand{\arraystretch}{1.12}
\caption{Classification accuracy (\%) on the DCASE 2024 Task~1 evaluation set across training-data fractions. Base rows (without “+”) use the official single-run accuracies reported by the DCASE Challenge; rows with “+” are our re-implementations reported as mean $\pm$ std over five independent experiments under a unified training setup.}

\label{tab:main_results}
\begin{tabular}{lccccc|cc}
\toprule
\textbf{System} & \textbf{5\%} & \textbf{10\%} & \textbf{25\%} & \textbf{50\%} & \textbf{100\%} & \textbf{Seen (5\%)} & \textbf{Unseen (5\%)} \\
\midrule
\textbf{DCASE2024 Baseline}~\cite{schmid2023distilling} 
& 44.00 & 46.95 & 51.47 & 54.40 & 56.84 & 45.30 & 42.40 \\
\quad +LCL          & 44.32 $\pm$ 1.23 & 46.20 $\pm$ 1.03 & 51.79 $\pm$ 0.86 & 54.40 $\pm$ 0.64 & 56.80 $\pm$ 0.44 & 45.25 $\pm$ 1.15 & 42.55 $\pm$ 1.39 \\
\quad +CLDG         & 44.89 $\pm$ 1.21 & 47.75 $\pm$ 0.99 & 52.14 $\pm$ 0.85 & 54.65 $\pm$ 0.62 & 56.80 $\pm$ 0.45 & 46.57 $\pm$ 1.12 & 43.12 $\pm$ 1.36 \\
\quad +EGCL         & 46.30 $\pm$ 1.19 & 49.10 $\pm$ 0.97 & 52.80 $\pm$ 0.81 & 55.20 $\pm$ 0.59 & 57.00 $\pm$ 0.41 & 47.50 $\pm$ 1.09 & 44.00 $\pm$ 1.33 \\
\quad +SSPL         & 46.98 $\pm$ 1.15 & 49.25 $\pm$ 0.96 & 53.85 $\pm$ 0.79 & 55.73 $\pm$ 0.58 & 57.58 $\pm$ 0.40 & 47.78 $\pm$ 1.11 & 44.54 $\pm$ 1.31 \\
\quad \textbf{+DDSC (ours)}  
                   & \textbf{48.17} $\pm$ 1.16 & \textbf{50.22} $\pm$ 0.94 & \textbf{54.31} $\pm$ 0.78 & \textbf{56.12} $\pm$ 0.57 & \textbf{58.19} $\pm$ 0.39 & \textbf{48.25} $\pm$ 1.07 & \textbf{46.10} $\pm$ 1.30 \\
\midrule
\textbf{Cai\_XJTLU}~\cite{cai2024dcase2024} 
& 48.91 & 53.16 & 58.09 & 59.47 & 62.12 & 50.60 & 46.70 \\
\quad +LCL          & 49.31 $\pm$ 1.23 & 53.55 $\pm$ 1.01 & 58.54 $\pm$ 0.84 & 60.00 $\pm$ 0.60 & 62.38 $\pm$ 0.42 & 50.92 $\pm$ 1.13 & 47.05 $\pm$ 1.37 \\
\quad +CLDG         & 49.86 $\pm$ 1.20 & 53.93 $\pm$ 1.03 & 58.92 $\pm$ 0.83 & 60.43 $\pm$ 0.59 & 62.74 $\pm$ 0.44 & 51.46 $\pm$ 1.12 & 47.72 $\pm$ 1.34 \\
\quad +EGCL         & 51.50 $\pm$ 1.18 & 55.10 $\pm$ 0.98 & 59.50 $\pm$ 0.82 & 61.25 $\pm$ 0.58 & 63.20 $\pm$ 0.40 & 52.30 $\pm$ 1.09 & 49.30 $\pm$ 1.31 \\
\quad +SSPL         & 51.95 $\pm$ 1.17 & 55.60 $\pm$ 0.99 & 60.10 $\pm$ 0.80 & 61.80 $\pm$ 0.57 & 63.64 $\pm$ 0.39 & 53.00 $\pm$ 1.08 & 50.05 $\pm$ 1.29 \\
\quad \textbf{+DDSC (ours)}  
                   & \textbf{53.70} $\pm$ 1.14 & \textbf{57.20} $\pm$ 0.95 & \textbf{61.75} $\pm$ 0.79 & \textbf{63.45} $\pm$ 0.56 & \textbf{64.25} $\pm$ 0.38 & \textbf{54.20} $\pm$ 1.06 & \textbf{51.68} $\pm$ 1.27 \\
\midrule
\textbf{DS-FlexiNet}~\cite{chen2025improving}
& 51.20 $\pm$ 1.36 & 55.10 $\pm$ 1.07 & 60.00 $\pm$ 0.89 & 62.10 $\pm$ 0.66 & 63.40 $\pm$ 0.49 & 52.36 $\pm$ 1.20 & 49.62 $\pm$ 1.53 \\
\quad +LCL          & 51.49 $\pm$ 1.27 & 55.44 $\pm$ 1.05 & 60.30 $\pm$ 0.86 & 62.40 $\pm$ 0.64 & 63.80 $\pm$ 0.46 & 52.64 $\pm$ 1.14 & 49.69 $\pm$ 1.41 \\
\quad +CLDG         & 51.92 $\pm$ 1.26 & 55.87 $\pm$ 1.03 & 60.60 $\pm$ 0.85 & 62.40 $\pm$ 0.62 & 63.78 $\pm$ 0.44 & 53.02 $\pm$ 1.16 & 50.34 $\pm$ 1.38 \\
\quad +EGCL         & 52.58 $\pm$ 1.19 & 56.42 $\pm$ 1.00 & 60.95 $\pm$ 0.82 & 62.92 $\pm$ 0.60 & 63.95 $\pm$ 0.42 & 53.60 $\pm$ 1.10 & 51.02 $\pm$ 1.33 \\
\quad +SSPL         & 54.08 $\pm$ 1.18 & 56.96 $\pm$ 0.99 & 61.40 $\pm$ 0.81 & 63.20 $\pm$ 0.59 & 64.18 $\pm$ 0.41 & 54.08 $\pm$ 1.12 & 51.62 $\pm$ 1.32 \\
\quad \textbf{+DDSC (ours)}  
                   & \textbf{55.56} $\pm$ 1.15 & \textbf{57.91} $\pm$ 0.96 & \textbf{61.95} $\pm$ 0.80 & \textbf{63.70} $\pm$ 0.58 & \textbf{64.84} $\pm$ 0.40 & \textbf{54.93} $\pm$ 1.09 & \textbf{52.75} $\pm$ 1.30 \\
\midrule
\textbf{Han\_SJTUTHU}~\cite{han2024data} 
& 54.35 & 56.69 & 59.09 & 60.38 & 61.82 & 55.60 & 52.70 \\
\quad +LCL          & 54.75 $\pm$ 1.21 & 56.05 $\pm$ 1.00 & 59.37 $\pm$ 0.82 & 60.62 $\pm$ 0.61 & 61.95 $\pm$ 0.41 & 55.82 $\pm$ 1.12 & 53.01 $\pm$ 1.33 \\
\quad +CLDG         & 55.25 $\pm$ 1.20 & 57.06 $\pm$ 1.00 & 59.83 $\pm$ 0.82 & 60.98 $\pm$ 0.60 & 62.08 $\pm$ 0.42 & 56.46 $\pm$ 1.12 & 53.68 $\pm$ 1.32 \\
\quad +EGCL         & 56.60 $\pm$ 1.17 & 58.25 $\pm$ 0.98 & 60.40 $\pm$ 0.79 & 61.65 $\pm$ 0.58 & 62.20 $\pm$ 0.41 & 57.30 $\pm$ 1.10 & 55.20 $\pm$ 1.29 \\
\quad +SSPL         & 57.22 $\pm$ 1.16 & 58.93 $\pm$ 0.97 & 61.05 $\pm$ 0.78 & 62.05 $\pm$ 0.57 & 62.48 $\pm$ 0.43 & 57.80 $\pm$ 1.09 & 55.75 $\pm$ 1.27 \\
\quad \textbf{+DDSC (ours)}  
                   & \textbf{57.86} $\pm$ 1.14 & \textbf{59.52} $\pm$ 0.96 & \textbf{61.68} $\pm$ 0.77 & \textbf{62.64} $\pm$ 0.56 & \textbf{63.03} $\pm$ 0.40 & \textbf{58.45} $\pm$ 1.07 & \textbf{56.42} $\pm$ 1.26 \\
\bottomrule
\end{tabular}
\end{table*}

\subsubsection{Evaluation Metrics}
Model performance is primarily evaluated using the official DCASE 2024 Task 1 metric: class-wise average accuracy. This metric computes the mean of per-class accuracies, ensuring a balanced performance assessment across all scene categories. It is defined as:
\begin{equation}
\text{Accuracy}_{\text{avg}} = \frac{1}{C} \sum_{c=1}^{C} \frac{N_c^{\text{correct}}}{N_c} \label{eq:accuracy}
\end{equation}
where \( C \) denotes the total number of scene classes, \( N_c \) represents the total number of samples belonging to class \( c \), and \( N_c^{\text{correct}} \) is the count of correctly classified samples for that specific class.

\subsubsection{Acoustic Scene Classification Baseline Systems}

To assess generality, we evaluate DDSC across several baselines from DCASE 2024 Task 1 Challenge and state-of-the-art models.

\begin{enumerate}
  \item \textbf{DCASE2024 Official Baseline}~\cite{schmid2023distilling}: uses CP-Mobile with knowledge distillation from PaSST and CP-ResNet, and applies Freq-MixStyle plus device impulse response simulation.

  \item \textbf{Cai\_XJTLU}~\cite{cai2024dcase2024} (ranked 4th): trains TF-SepNet-64 via distillation from a pretrained PaSST teacher and employs SpecAugment, Mixup, additive noise, and pseudo-labeling.

  \item \textbf{Han\_SJTUTHU}~\cite{han2024data} (ranked 1st): distills from a PaSST ensemble to SSCP-Mobile and further uses SpecAugment, Freq-MixStyle, and pruning.

  \item \textbf{DS-FlexiNet}~\cite{chen2025improving}: combines a depthwise-separable CNN with residual connections, distills from multiple teachers, adopts ADIR and Freq-MixStyle, and supports int8 quantization-aware training.
\end{enumerate}

\subsection{Compared Training Strategies}
\label{subsec:training_strategies}

We compare DDSC with the following training strategies.

\begin{enumerate}
  \item \textbf{Entropy-Guided Curriculum (EGCL)}~\cite{Zhang2025}: ranks samples by device-posterior entropy and trains from high-entropy (domain-invariant) to low-entropy (domain-specific).

  \item \textbf{Symmetric Self-Paced Learning (SSPL)}~\cite{zhao2024symmetric}: uses a symmetric self-paced scheduler with a gradient-based difficulty measure to shift weights from easy to hard examples over epochs for domain generalization.

  \item \textbf{Ladder Curriculum Learning (LCL)}~\cite{wang2024ladder}: stages training by ordering data from easy to hard at both inter-domain and intra-domain levels for better domain generalization.

  \item \textbf{Curriculum Learning-based Domain Generalization (CLDG)} \cite{wang2024curriculum}: uses a mixup-based reciprocal-point classifier and a conditional domain discriminator, and orders training by category observation degree to handle domain shift.

\end{enumerate}

\subsection{Results and Analysis}

As shown in Table~\ref{tab:main_results}, DDSC improves accuracy across all backbones and label fractions, with the strongest gains in the data-efficient regime. At the 5\% label budget, training with DDSC increases overall accuracy by about 4.2\% on average across the four systems and raises unseen-device accuracy by about 3.9\%, indicating stronger cross-device generalization under scarce supervision. For each backbone and budget, models trained with DDSC achieve the best (or tied-best) results, improving robustness to unseen devices without degrading seen-device performance. These results show that the proposed dual-signal, time-varying schedule enhances data efficiency without architectural changes or inference overhead.


\section{Conclusion}
We introduced the Dynamic Dual-Signal Curriculum (DDSC), a dynamic training schedule for Acoustic Scene Classification (ASC) under domain shift. DDSC combines a domain-invariance signal derived from prototype–posterior entropy with a learning-progress signal, and fuses them via a time-varying scheduler into per-example weights. Under the DCASE 2024 Task~1 protocol, DDSC consistently improves accuracy across diverse backbones and label budgets—most notably in low-label settings—while requiring no architectural changes or inference overhead. DDSC is complementary to data augmentation and feature-alignment methods; future work will extend it beyond ASC and relax the need for device labels via pseudo-device discovery.

\section{Acknowledgment}
This work was funded by Basic Research Program of Jiangsu Province under Grant BG2024027.

\newpage

\bibliographystyle{IEEEbib}
\bibliography{refs}

\end{document}